\documentstyle[11pt,epsf,aaspp4]{article}
\def\zmax{{z_{\rm max}}}
\def\zmin{{z_{\rm min}}}
\def\pp{\par\parshape 2 0truecm 15.5truecm 1truecm 14.5truecm\noindent}
\newcommand{\simgt}{\lower.5ex\hbox{$\; \buildrel > \over \sim \;$}}
\newcommand{\simlt}{\lower.5ex\hbox{$\; \buildrel < \over \sim \;$}}
\newcommand{\himpc}{{\hbox {$h^{-1}$}{\rm Mpc}} }
\newcommand{\PS}{ {\scriptscriptstyle {\rm PS}} }

\newcommand{\A}{{\scriptscriptstyle A}}
\newcommand{\B}{{\scriptscriptstyle B}}
\newcommand{\C}{{\scriptscriptstyle C}}
\renewcommand{\L}{{\scriptscriptstyle L}}
\newcommand{\bol}{{\rm\scriptscriptstyle bol}}
\newcommand{\gas}{{\rm\scriptscriptstyle gas}}
\newcommand{\band}{{\rm\scriptscriptstyle band}}
\begin{document}

\begin{minipage}[c]{4cm}
RESCEU-24/99\\
UTAP-334/99\\
HUPD-9908\\
astro-ph/9907105
\end{minipage}\\

\title{
Two-point correlation functions of X-ray selected clusters of galaxies: \\
  theoretical predictions for flux-limited surveys
}

\bigskip

\author{
  Yasushi Suto\\ 
{\it Department of Physics and Research Center for
    the Early Universe (RESCEU) \\ School of Science, University of
    Tokyo, Tokyo 113-0033, Japan. } \\
Kazuhiro Yamamoto \\ {\it Department of Physics, Hiroshima
    University, Higashi-Hiroshima 739-8526, Japan.} \\
  Tetsu Kitayama\\ 
{\it Department of Physics, Tokyo Metropolitan University,
    Hachioji 192-0397, Japan. } 
\\ and \\ 
  Y.P.Jing\\ 
{\it Research Center for
    the Early Universe (RESCEU) \\ School of Science, University of
    Tokyo, Tokyo 113-0033, Japan. } 
}

\bigskip

\affil{ e-mail: suto@phys.s.u-tokyo.ac.jp, 
yamamoto@astro.phys.sci.hiroshima-u.ac.jp, 
tkita@phys.metro-u.ac.jp,
jing@utaphp1.phys.s.u-tokyo.ac.jp
}

\received{1999 July 8}
\accepted{1999 November 23}

\begin{abstract}
We have developed a model to describe two-point correlation functions of
clusters of galaxies in X-ray flux-limited surveys. Our model properly
takes account of nonlinear gravitational evolution of mass fluctuations,
redshift-space distortion due to linear peculiar velocity field and to
finger-of-god, cluster abundance and bias evolution on the basis of the
Press -- Schechter theory, the light-cone effect, and the selection
function due to the X-ray flux, temperature and luminosity
limits. Applying this model in representative cosmological models, we
have presented quantitative predictions for X-ray selected samples
feasible from the future surveys with the X-ray satellites including
Astro-E, Chandra, and XMM. The comparison of these predictions and the
observed cluster clustering will place important cosmological
constraints which are complementary to the cluster abundance and the
cosmic microwave background.
\end{abstract}


%%%%%%%%%%%%%%%%%%%%%%%%%%%%%%%%%%%%%%%%%%%%%%%%%%%%%%%%%%%%%%%%%%%%%%%
\clearpage

\baselineskip 15pt

%%%%%%%%%%%%%%%%%%%%%%%
\section{Introduction}
%%%%%%%%%%%%%%%%%%%%%%%

The abundance of cluster of galaxies is now well-established as a
standard cosmological probe (White, Efstathiou \& Frenk 1991; Jing \&
Fang 1994; Barbosa et al. 1996; Viana \& Liddle 1996; Eke, Cole, \&
Frenk 1996; Kitayama \& Suto 1997). In particular, several available
X-ray catalogues of clusters have played an important role in placing
fairly robust constraints on the cosmological parameters.  The spatial
two-point correlation function of clusters is another important target
for cosmological researches (Bahcall \& Soneira 1983; Klypin \&
Kopylov 1983; Bahcall 1988; Bahcall \& Cen 1993; Ueda, Itoh \& Suto
1993; Jing et al. 1993; Watanabe, Matsubara \& Suto 1994; Borgani,
Plionis, \& Kolokotronis 1999). In fact the idea of biased galaxy
formation by Kaiser (1984) was proposed originally to reconcile the
stronger spatial correlation of clusters with those of galaxies.  The
previous studies in this context, however, have been mainly based on
the optically selected samples, which are likely to be contaminated by
the projection effect and thus the selection function of which is
difficult to evaluate precisely.

The X-ray flux-limited catalogues of clusters are now becoming
available with ROSAT (e.g. Ebeling et al. 1997; Rosati et al. 1998)
and are expected to increase in their sample volume in near future
with the X-ray satellites including Astro-E, Chandra, and XMM. Such
well-controlled catalogues are ideal to revisit the cluster
correlation functions with unprecedented precision. The proper
comparison with such data, however, requires better theoretical
predictions which take account of the selection function of X-ray
clusters (Kitayama, Sasaki \& Suto 1998), the luminosity and time
dependent bias (Mo \& White 1996; Jing 1998; Moscardini et al. 1998),
the light-cone effect (Matarrese et al. 1997; Matsubara, Suto \&
Szapudi 1997; Nakamura, Matsubara \& Suto 1998; Yamamoto \& Suto 1999;
Moscardini et al. 1999a) and the redshift-space distortion (Hamilton
1998; Matsubara \& Suto 1996; Suto et al. 1999; Nishioka \& Yamamoto
1999; Yamamoto, Nishioka \& Suto 1999; Magira, Jing \& Suto 2000). The
theoretical prescriptions for those effects have been developed and
become available recently. This motivates us to present detailed
predictions for the two-point correlation functions for X-ray
flux-limited samples of clusters of galaxies combining the cluster
abundances in a fully consistent fashion.

Moscardini et al. (1999a) recently performed a similar work, and our
present paper differs from theirs in several aspects. First, we take
account of the linear and nonlinear redshift-space distortion due to
the peculiar velocity field.  Second we adopt a formula for the
light-cone effect derived by Yamamoto \& Suto (1999) which is the
one-dimensional integration over the redshift in the cluster sample.
Finally, we are mainly interested in future surveys which probe a
higher redshift ($z \sim 1$) universe, and therefore focus on a range
of the X-ray flux limit a few magnitudes fainter than they considered.

Koyama, Soda \& Taruya (1999) suggested a presence of primordial
non-Gaussianity from the combined analysis of the cosmic microwave
background, the abundance of X-ray clusters at $z=0$ and $0.3$, and
the correlation length of optical cluster samples. While this
interpretation is interesting, the definite conclusion requires the
proper account of the selection effect as well as the theoretical
modeling and the statistical limitation of the available sample. In
this sense, optically selected cluster samples currently available are
still far from satisfactory, and the analysis of the upcoming X-ray
selected catalogues is essential.

%%%%%%%%%%%%%%%%%%%%%%%%%%%%%%%%%%%%%%%%%%%%%%%%%%%%%%%%%%%%%%%%%%%%%%%%%%
\section{Modelling the two-point correlation functions of X-ray clusters}
%%%%%%%%%%%%%%%%%%%%%%%%%%%%%%%%%%%%%%%%%%%%%%%%%%%%%%%%%%%%%%%%%%%%%%%%%%

%%%%%%%%%%%%%%%%%%%%%%%%%%%%%%%%%%%%%%%%%%%%%%%%%%%%%%%%%%%%%%
\subsection{Linear and nonlinear redshift-space distortion}
%%%%%%%%%%%%%%%%%%%%%%%%%%%%%%%%%%%%%%%%%%%%%%%%%%%%%%%%%%%%%%

The observable two-point correlation functions are inevitably distorted
due to the presence of the peculiar velocity field. We take into account
this redshift-space distortion following Cole et al. (1994), Magira et
al.(2000) and Yamamoto et al. (1999).

Our key assumption is that the bias of the cluster density field relative
to the mass density field is linear and scale-independent:
%%%%%%%%%%%%%%%%%%%%%%%%%%%%%%%%%%%%%%%%%%%%%%%%%%%%%%%%%%%%%%%%%%%%%
\begin{eqnarray}
\label{eq:biasdef}
 \delta_{\rm cl}({\bf x},z)  = 
     b_{\rm cl}(z)~\delta_{\rm mass}({\bf x},z) .
\end{eqnarray}
%%%%%%%%%%%%%%%%%%%%%%%%%%%%%%%%%%%%%%%%%%%%%%%%%%%%%%%%%%%%%%%%%%%%%
In this case, the power spectrum of the corresponding cluster samples in
redshift space is well approximated as
%%%%%%%%%%%%%%%%%%%%%%%%%%%%%%%%%%%%%%%%%%%%%%%%%%%%%%%%%%%%%%%%%%%%%%
\begin{eqnarray}
  P^{\rm S}_{\rm cl}(k,\mu,z)
=\left[ {1+\beta(z)\mu^2 \over 1+(k\mu\sigma_v)^2/2}  \right]^2
 \left[b_{\rm cl}(z)\right]^2~ P^{\rm R}_{\rm mass}(k,z) ,
\label{eq:Pkcl2d}
\end{eqnarray}
%%%%%%%%%%%%%%%%%%%%%%%%%%%%%%%%%%%%%%%%%%%%%%%%%%%%%%%%%%%%%%%%%%%%%%
where $\mu$ the direction cosine of the wavenumber vector and the line-of-sight
of the fiducial observer, and $P^{\rm R}_{\rm mass}(k,z)$ is the mass power
spectrum in real space.  The numerator in equation (\ref{eq:Pkcl2d})
expresses the linear redshift-space distortion (Kaiser 1987), where
$\beta$ is defined by
%%%%%%%%%%%%%%%%%%%%%%%%%%%%%%%%%%%%%%%%%%%%%%%%%%%%%%%%%%%%%%%
\begin{eqnarray}
  \beta(z)={1\over b_{\rm cl}(z)} {d\ln D_1(z)\over d\ln a(z)},
\label{defbeta}
\end{eqnarray}
%%%%%%%%%%%%%%%%%%%%%%%%%%%%%%%%%%%%%%%%%%%%%%%%%%%%%%%%%%%%%%%%
and $D_1(z)$ is the linear growth factor normalized to be unity at
present. The denominator in equation (\ref{eq:Pkcl2d}) takes account of
the nonlinear redshift-space distortion ({\it finger-of-God}) assuming
that the one-point distribution function of the peculiar velocity is
exponential with the velocity dispersion of $\sigma_v(z)$.

Averaging equation (\ref{eq:Pkcl2d}) over the angle with respect to
the line-of-sight of the observer yields
%%%%%%%%%%%%%%%%%%%%%%%%%%%%%%%%%%%%%%%%%%%%%%%%%%%%%%%%%%%%%%%%%%%%%%%%%%
\begin{eqnarray}
\label{eq:Pkcl0}
  P^{\rm S}_{\rm cl}(k,z) &=&
\left[A(\kappa)+{2\over3}\beta(z) B(\kappa)+{1\over 5}\beta^2(z) 
C(\kappa)\right]   \left[b_{\rm cl}(z)\right]^2~ P^{\rm R}_{\rm mass}(k,z) 
\\
A(\kappa) &=& {{\rm arctan}(\kappa/\sqrt{2})\over \sqrt{2}\kappa}+
  {1\over 2+\kappa^2},
\\
  B(\kappa) &=& {6\over\kappa^2}\biggl(A(\kappa)-{2\over 2+\kappa^2}\biggr),
\\
  C(\kappa) &=& {-10\over\kappa^2}\biggl(B(\kappa)-{2\over 2+\kappa^2}\biggr),
\end{eqnarray}
%%%%%%%%%%%%%%%%%%%%%%%%%%%%%%%%%%%%%%%%%%%%%%%%%%%%%%%%%%%%%%%%%%
with $\kappa(z)=k\sigma_v(z)/H_0$. Finally the corresponding
two-point correlation function of clusters in redshift space is
computed as
%%%%%%%%%%%%%%%%%%%%%%%%%%%%%%%%%%%%%%%%%%%%%%%%%%%%%%%%%%%%%%%%%%%
\begin{equation}
\label{eq:xiScl}
  \xi^{\rm S}_{\rm cl}(R,z) = {1\over2\pi^2}
   \int_0^\infty  dk k^2 P^{\rm S}_{\rm cl}(k,z) j_0(kR) ,
\end{equation}
%%%%%%%%%%%%%%%%%%%%%%%%%%%%%%%%%%%%%%%%%%%%%%%%%%%%%%%%%%%%%%%%%%%
where $j_0(kR)$ is the spherical Bessel function.  In what follows, we
adopt nonlinear evolution of the mass fluctuations using the Peacock
\& Dodds (1996; PD) fitting formula for $P^{\rm R}_{\rm mass}(k,z)$
unless otherwise stated.

%%%%%%%%%%%%%%%%%%%%%%%%%%%%%%%%%%%%%%%%%%%%%%%%%%%%%%%%%%%
\subsection{Temperature and luminosity of X-ray clusters}
%%%%%%%%%%%%%%%%%%%%%%%%%%%%%%%%%%%%%%%%%%%%%%%%%%%%%%%%%%%

The X-ray luminosity function of galaxy clusters at $z\sim 0$ is
determined to a good precision from existing observational catalogues
(Burns et al. 1996; Ebeling et al. 1997). In fact, the Press-Schechter
theory applied to the CDM models reproduces quite well the X-ray
luminosity function, as well as the X-ray temperature function,
provided that the amplitude of the mass fluctuation at $8\himpc$,
$\sigma_8$, is related to the density parameter $\Omega_0$ and the
cosmological constant $\lambda_0$ as follows (Kitayama \& Suto 1997;
see also Viana \& Liddle 1996; Eke, Cole \& Frenk 1996):
%%%%%%%%%%%%%%%%%%%%%%%%%%%%%%%%%%%%%%%%%%%%%%%%%%%%%%%%%%%%%%%%%%%
\begin{eqnarray}
\label{eq:s8o0}
  \sigma_8 = 0.54  \times \left\{
      \begin{array}{ll}
        \Omega_0^{-0.35-0.82\Omega_0+0.55\Omega_0^2} &
        \mbox{($\lambda_0=1-\Omega_0$)}, \\ 
        \Omega_0^{-0.28-0.91\Omega_0+0.68\Omega_0^2} &
        \mbox{($\lambda_0=0$)} 
      \end{array}
   \right. .
\end{eqnarray}
%%%%%%%%%%%%%%%%%%%%%%%%%%%%%%%%%%%%%%%%%%%%%%%%%%%%%%%%%%%%%%%%%%%

In this paper, we follow the prescription of Kitayama \& Suto (1997)
to relate the mass $M$ of the ``Press-Schechter objects'' to the X-ray
luminosity $L$ (bolometric or band-limited) of the real clusters of
galaxies. Although the one-to-one correspondence of those two species
is a non-trivial assumption, it is regarded as a fairly successful
approximation, at least empirically.  We first relate the total mass
$M$ of the dark halo of a cluster to the temperature of hot gas,
$T_{\gas}$, assuming the virial equilibrium:
%%%%%%%%%%%%%%%%%%%%%%%%%%%%%%%%%%%%%%%%%%%%%%%%%%%%%%%%%%%%%%
\begin{eqnarray}
  k_\B T_{\gas} &=& \gamma {\mu m_p G M \over 3 r_{\rm vir}(M,z)},
  \nonumber \\ &=& 5.2\gamma (1+z) \left({\Delta_{\rm vir} \over
      18\pi^2}\right)^{1/3} \left({M \over 10^{15} h^{-1} M_\odot}
  \right)^{2/3} \Omega_0^{1/3} ~{\rm keV}.
\label{eq:tm}
\end{eqnarray}
%%%%%%%%%%%%%%%%%%%%%%%%%%%%%%%%%%%%%%%%%%%%%%%%%%%%%%%%%%%%%%   
where $k_\B$ is the Boltzmann constant, $G$ is the gravitational
constant, $m_p$ is the proton mass, $\mu$ is the mean molecular weight
(we adopt $\mu=0.59$), and $\gamma$ is a fudge factor of order unity
(we adopt $\gamma=1.2$). In the above, $z$ is the redshift of the
cluster which we assume is equal to the cluster formation epoch for
simplicity (see Kitayama \& Suto 1996 for more discussion on this
point). While the cluster temperature may not be isothermal, it is not
important in our analysis. In fact, the above temperature should be
interpreted as an average temperature of the cluster. In fact, our
latest hydrodynamical simulations (Yoshikawa, Jing \& Suto 1999) show
that the mass- and emission-weighted temperatures of simulated
clusters satisfy the above relation with $\gamma=1.2$ and $1.6$,
respectively. Further details of the temperature-mass relation and its
non-isothermal effect are discussed in Makino, Sasaki \& Suto (1998),
Yoshikawa, Itoh, \& Suto (1998), Suto, Sasaki \& Makino (1999) and
Yoshikawa \& Suto (1999).  We compute $\Delta_{\rm vir}$, the ratio of
the mean cluster density to the mean density of the universe at that
epoch using the formulae for the spherical collapse model presented in
Kitayama \& Suto (1996).

 Next we transform the temperature to the luminosity of clusters using
the {\it observed} luminosity -- temperature relation:
%%%%%%%%%%%%%%%%%%%%%%%%%%%%%%%%%%%%%%%%%%%%%%%%%%%%%%%%%%%%%%
\begin{equation}
  L_\bol = L_{44} \left( \frac{T_{\gas}}{6{\rm keV}} 
\right)^{\alpha}
  (1+z)^\zeta ~~ 10^{44} h^{-2}{\rm ~ erg~sec}^{-1} .
\label{eq:lt}
\end{equation}
%%%%%%%%%%%%%%%%%%%%%%%%%%%%%%%%%%%%%%%%%%%%%%%%%%%%%%%%%%%%%%
While a simple theory on the basis of the self-similar cluster
evolution predicts the slope $\alpha=2$, we adopt $L_{44}=2.9$,
$\alpha=3.4$ and $\zeta=0$ on the basis of recent observational
indications (e.g., David et al. 1993; Ebeling et al. 1996; Ponman et
al. 1996; Mushotzky \& Scharf 1997).  Then we translate
$L_\bol(T_{\gas})$ into the band-limited luminosity
$L_\band[T_{\gas},E_1,E_2]$ as
%%%%%%%%%%%%%%%%%%%%%%%%%%%%%%%%%%%%%%%%%%%%%%%%%%%%%%%%%%%%%%
\begin{eqnarray}
  L_\band[T_{\gas},E_a(1+z),E_b(1+z)] = L_\bol (T_{\gas})\times
  f[T_{\gas},E_a(1+z),E_b(1+z)],
\label{eq:band}
\end{eqnarray}
%%%%%%%%%%%%%%%%%%%%%%%%%%%%%%%%%%%%%%%%%%%%%%%%%%%%%%%%%%%%%%
where $f[T_{\gas},E_1,E_2]$ is the band correction factor which takes
account of metal line emissions (Masai 1984) in addition to the
thermal bremsstrahlung.

Finally the source luminosity $L_\band$ at $z$ is converted to 
the observed flux $S$ in an X-ray energy band [$E_a$,$E_b$]:
%%%%%%%%%%%%%%%%%%%%%%%%%%%%%%%%%%%%%%%%%%%%%%%%%%%%%%%%%%%%%%
\begin{equation}
S[E_a,E_b] = {L_\band[E_a(1+z),E_b(1+z)] \over 4 \pi d_\L^2(z) }
\label{eq:ls}  
\end{equation}
%%%%%%%%%%%%%%%%%%%%%%%%%%%%%%%%%%%%%%%%%%%%%%%%%%%%%%%%%%%%%%
where $d_\L(z)$ is the luminosity distance.  Throughout the present
paper, we use the $0.5-2.0$keV band for the flux assuming the
abundance of intracluster gas as 0.3 times the solar value.  The
luminosity distance is related to the comoving distance $d_\C$ and the
angular diameter distance $d_\A$ as $d_\L(z) = (1+z) d_\C(z) = (1+z)^2
d_\A(z)$;
%%%%%%%%%%%%%%%%%%%%%%%%%%%%%%%%%%%%%%%%%%%%%%%%%%%%%%%%%%%%%%%%%%%
\begin{eqnarray}
 d_\C(z) &=& 
  \left\{ 
      \begin{array}{ll}
         H_0^{-1} \sin{(H_0 r \sqrt{\Omega_0 + \lambda_0 -1} )}
          /\sqrt{\Omega_0 + \lambda_0 - 1} 
              & \mbox{$(\Omega_0 + \lambda_0 >1)$} \\
         r & \mbox{$(\Omega_0 + \lambda_0 =1)$} \\
         H_0^{-1} \sinh{(H_0 r \sqrt{1-\Omega_0-\lambda_0} )}
          /\sqrt{1-\Omega_0-\lambda_0} 
         & \mbox{$(\Omega_0 + \lambda_0 <1)$} 
      \end{array}
   \right. ,
\end{eqnarray}
%%%%%%%%%%%%%%%%%%%%%%%%%%%%%%%%%%%%%%%%%%%%%%%%%%%%%%%%%%%%%%%%%%%
where $r(z)$ is the radial distance:
%%%%%%%%%%%%%%%%%%%%%%%%%%%%%%%%%%%%%%%%%%%%%%%%%%%%%%%%%%%%%%%%%%%
\begin{equation}
\label{eq:rz}
r(z) = \int_0^z {dz \over H(z)} ,
\end{equation}
%%%%%%%%%%%%%%%%%%%%%%%%%%%%%%%%%%%%%%%%%%%%%%%%%%%%%%%%%%%%%%%%%%%
and $H(z)$ the Hubble parameter at redshift $z$:
%%%%%%%%%%%%%%%%%%%%%%%%%%%%%%%%%%%%%%%%%%%%%%%%%%%%%%%%%%%%%%%%%%%
\begin{equation}
\label{eq:hz}
   H(z) = H_0\sqrt{\Omega_0 (1 + z)^3 +
(1-\Omega_0-\lambda_0) (1 + z)^2 + \lambda_0}  .
\end{equation}
%%%%%%%%%%%%%%%%%%%%%%%%%%%%%%%%%%%%%%%%%%%%%%%%%%%%%%%%%%%%%%%%%%%

In what follows, we mainly consider three representative models; SCDM
(Standard CDM) with $(\Omega_0,\lambda_0,h,\sigma_8)$
$=(1.0,0.0,0.5,0.56)$, LCDM (Lambda CDM) with $(0.3,0.7,0.7,1.04)$, and
OCDM (Open CDM) with $(0.45,0.0,0.7,0.83)$, where $h$ is the Hubble
constant $H_0$ in units of $100$km/s/Mpc.  It should be noted that
equation (\ref{eq:s8o0}) is valid for $h=0.7$, strictly speaking. For
that reason, we adopt $\sigma_8=0.56$, instead of $0.54$, to match the
cluster abundance for SCDM with $h=0.5$.

\begin{figure}
\begin{center}
    \leavevmode\epsfysize=10.0cm \epsfbox{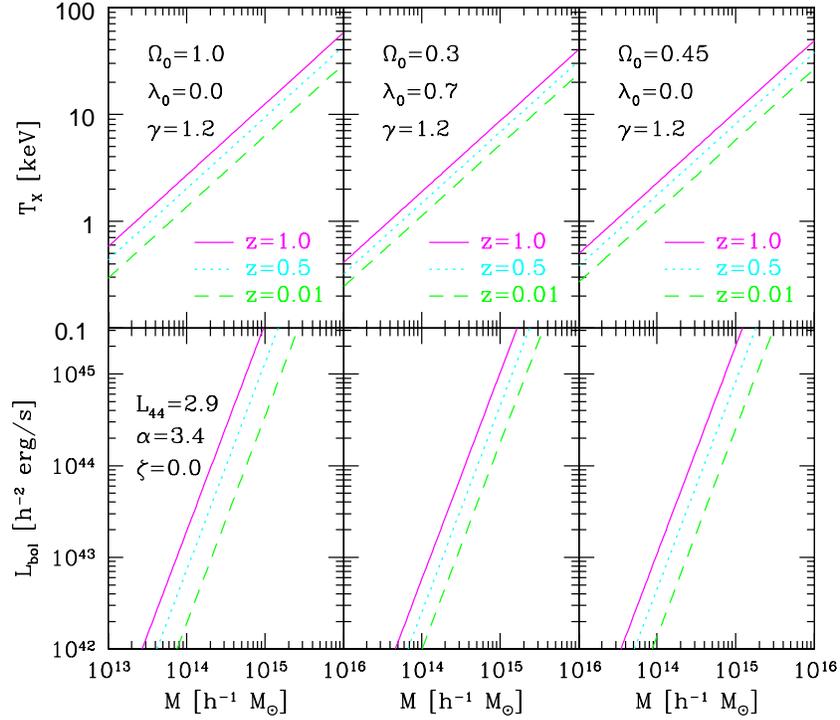}
\end{center}
\figcaption{The X-ray temperature and luminosity as a function of
  cluster mass for three representative sets of cosmological
  parameters; $(\Omega_0,\lambda_0)=(1.0,0.0)$, $(0.3,0.7)$, and
  $(0.45,0.0)$ from left to right panels.  The X-ray temperature ({\it
  Upper panels}) and the bolometric X-ray luminosity ({\it Lower
  panels}) are plotted at $z=1.0$ ({\it solid lines}), $0.5$ ({\it
  dotted lines}) and $0.01$ ({\it dashed lines}). We adopt the ratio
  of gas to virial temperature $\gamma=1.2$
  (eq.[\protect\ref{eq:tm}]), and the non-evolving ($\zeta=0$)
  luminosity--temperature relation (eq.[\protect\ref{eq:lt}]) with the
  amplitude $L_{44}=2.9$ and the power-law index $\alpha=3.4$.
\label{fig:tlm}}
\end{figure}

\begin{figure}
\begin{center}
    \leavevmode\epsfysize=10.0cm \epsfbox{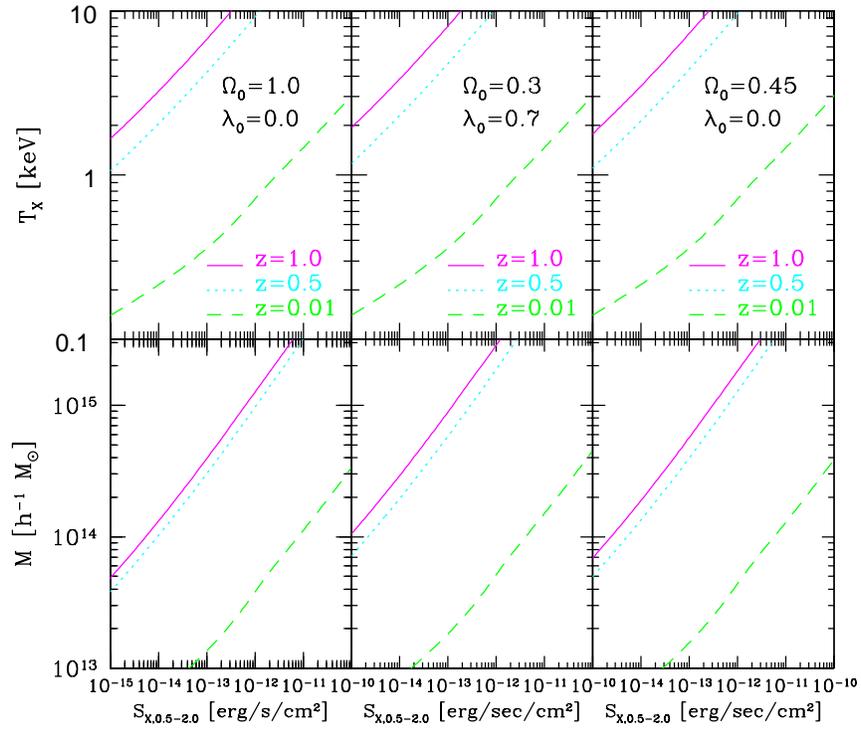}
\end{center}
\figcaption{The X-ray temperature and cluster mass as a function of X-ray
  flux in the $0.5 - 2$ keV band.  Different lines indicate results at
  different redshifts; $z=1.0$ ({\it solid}), $0.5$ ({\it dotted}) and
  $0.01$ ({\it dashed}). \label{fig:tmsx}}
\end{figure}

\begin{figure}
\begin{center}
    \leavevmode\epsfxsize=12cm \epsfbox{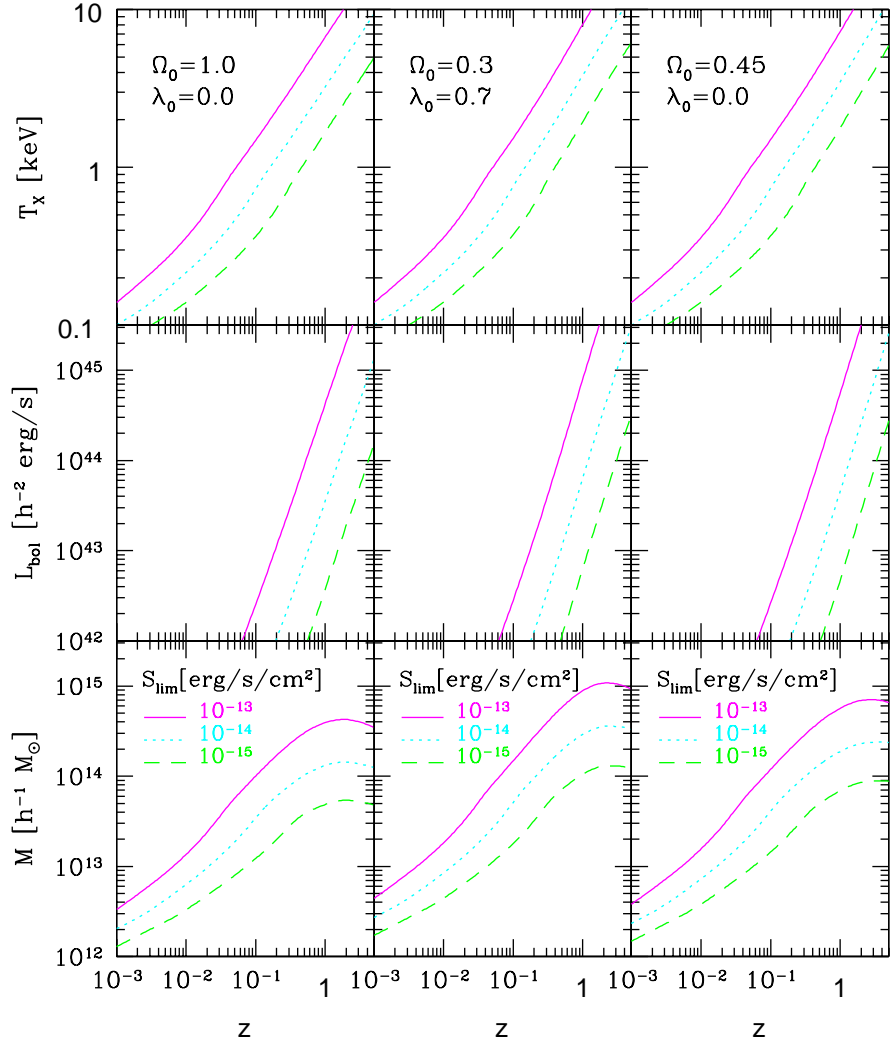}
\end{center}
\figcaption{ Temperature, bolometric luminosity and mass of clusters
corresponding to the X-ray flux-limit $S_{\rm lim}$ as a function of
redshift $z$; $S_{\rm lim}=10^{-13}$ ({\it solid lines}), $10^{-14}$
({\it dotted}) and $10^{-15}$erg/s/cm$^2$ ({\it dashed}).
\label{fig:tlmz}}
\end{figure}

The relations among mass, temperature, luminosity and flux of X-ray
clusters which we describe in the above do not depend on the power
spectrum. For definiteness, however, we choose the same cosmological
parameters as the above three models, and plot the results.  Figure
\ref{fig:tlm} shows the temperature $T_{\rm X}$ and the bolometric
luminosity $L_{\rm bol}$, as a function of cluster mass $M$ at
$z=0.01$ (dashed lines), 0.5 (dotted lines) and 1.0 (solid
lines). Figure \ref{fig:tmsx} plots $T_{\rm X}$ and $M$ as a function
of X-ray flux $S_{0.5-2.0}$ for clusters at $z=0.01$ (dashed lines),
0.5 (dotted lines) and 1.0 (solid lines).  Figure \ref{fig:tlmz} shows
$T_{\rm X}$, $L_{\rm bol}$ and $M$ of clusters at $z$ with the
observed X-ray flux $S_{0.5-2.0} = 10^{-13}$, $10^{-14}$ and
$10^{-15}$erg/s/cm$^2$ in solid, dotted and dashed curves,
respectively.

%%%%%%%%%%%%%%%%%%%%%%%%%%%%%%%%%%%%%%%%%%%%%%%%%%%%%%
\subsection{Evolution of luminosity--dependent bias}
%%%%%%%%%%%%%%%%%%%%%%%%%%%%%%%%%%%%%%%%%%%%%%%%%%%%%%

In order to complete a model for $\xi^{\rm S}_{\rm cl}(R,z;>S_{\rm
lim})$, the two-point correlation function of X-ray clusters on a
constant-time hypersurface at $z$, one has to specify a model for bias.
Here we adopt a fitting formula for bias of Jing (1998), which improves
an original proposal by Mo \& White (1996) on the basis of
high-resolution N-body simulations:
%%%%%%%%%%%%%%%%%%%%%%%%%%%%%%%%%%%%%%%%%%%%%%%%%%%%%%%%%%%%%%%%%%%%%
\begin{equation}
\label{eq:biasjing}
b(z,M)=\left[{0.5 \delta_c^4\over \Delta^4(M,z)}+1 
\right]^{0.06-0.02n_{\rm eff}} 
\left[1 - {1\over\delta_c} + {\delta_c\over\Delta^2(M,z)}\right] .
\end{equation}
%%%%%%%%%%%%%%%%%%%%%%%%%%%%%%%%%%%%%%%%%%%%%%%%%%%%%%%%%%%%%%%%%%%%%
In the above, $\Delta(M,z)$ is the density fluctuation smoothed over a
(top-hat) mass scale of $M$ at the redshift $z$, $n_{eff}$ is the
effective slope of the linear density power spectrum at the mass scale
$M$ (see Jing 1998), $\delta_c=1.69$ is the critical density contrast
for the spherical collapse.  To be specific, Jing (1998) carried out
several simulations in four scale-free models and three representative
CDM models employing $256^3$ particles. Each model is simulated with
three or four different realizations. The dark matter halos are
identified using the friends-of-friends algorithm using the linking
length of 0.2 times the mean particle separation. Thus the above
formula (\ref{eq:biasjing}) applies for the virialized dark halo
defined according to the Press -- Schechter theory, and is appropriate
for the clusters of galaxies. 

Combining the results in the previous subsection, we translate the above
bias factor into a function of X-ray flux limit according to
%%%%%%%%%%%%%%%%%%%%%%%%%%%%%%%%%%%%%%%%%%%%%%%%%%%%%%%%%%%%%%%%%%%%%
\begin{eqnarray}
\label{eq:beff}
 b_{\rm eff}(z, >S_{\rm lim}) &=&
   {\displaystyle \int_{M_{\rm lim}(S_{\rm lim})}^\infty 
dM ~b(z,M)~ n_\PS(z,M)
\over
\displaystyle \int_{M_{\rm lim}}^\infty dM ~ n_\PS(z,M)} \\
\label{eq:psmass}
n_\PS(z,M) &=& - \left( {2 \over \pi } \right)^{1/2} 
        {3\Omega_0H_0^2 \over 8\pi G M} {\delta_c \over \Delta^2(M,z)} 
        {d \Delta(M,z) \over d M} 
    \exp \left[ - { \delta_c^2 \over 2 \Delta^2(M,z)} \right] , 
\end{eqnarray}
%%%%%%%%%%%%%%%%%%%%%%%%%%%%%%%%%%%%%%%%%%%%%%%%%%%%%%%%%%%%%%%%%%%%%
where $n_\PS(z,M)$ is the Press -- Schechter mass function.  Given the
flux limit $S_{\rm lim}$, the corresponding luminosity $L_{\rm lim}(z)$
for a cluster located at $z$ is computed from equation (\ref{eq:ls})
with the luminosity distance $d_{\rm L}(z)$. Then we solve equations
(\ref{eq:lt}) and (\ref{eq:band}) for $T_{\rm lim}(z) =T_{\gas}$, and
finally obtain $M_{\rm lim}(z)$ using equation (\ref{eq:tm}).

\begin{figure}
\begin{center}
    \leavevmode\epsfxsize=12cm \epsfbox{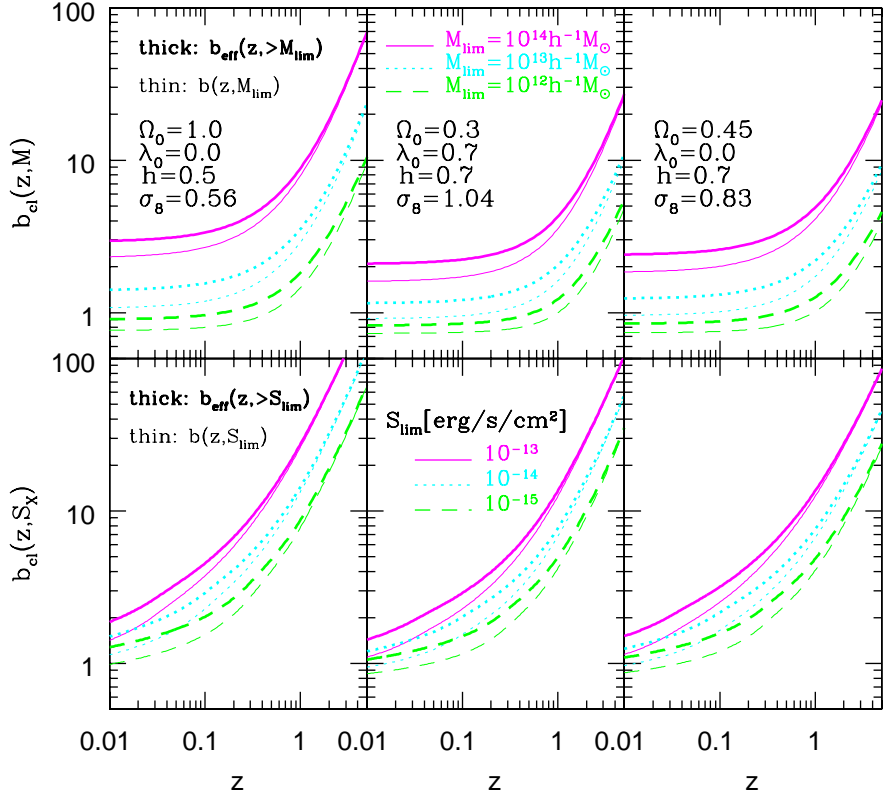}
\end{center}
\figcaption{Evolution of bias for clusters selected by mass and X-ray
flux in SCDM, LCDM and OCDM models; {\it Upper panels:} bias for
different mass limits, $M_{\rm lim}=10^{14}h^{-1}M_\odot$,
$10^{13}h^{-1}M_\odot$ and $10^{12}h^{-1}M_\odot$ in solid, dotted and
dashed lines, respectively.  Thin lines indicate the result for
$b(z,M_{\rm lim})$ while thick lines for the effective bias $b_{\rm
eff}(z,>M_{\rm lim})$.  {\it Lower panels:} bias for different flux
limits, $S_{\rm lim}=10^{-13}$, $10^{-14}$ and $10^{-15}$erg/s/cm$^2$
in solid, dotted and dashed lines, respectively.  Thin lines indicate
the result for $b(z,S_{\rm lim})$ while thick lines for the effective
bias $b_{\rm eff}(z,>S_{\rm lim})$.  \label{fig:biasmsx}}
\end{figure}

Figure \ref{fig:biasmsx} plots the evolution of $b_{\rm eff}(z,>M)$,
$b(z,M)$, $b_{\rm eff}(z,>S_{\rm lim})$ and $b(z,S_{\rm lim})$ for our
model of X-ray clusters.  As noted in Figure 3 of Jing (1998), the
fitting formula is accurate within 10 percent for $M \simlt
10^{14}h^{-1}M_\odot$ on linear scales. We checked the accuracy of the
fit using the more recent numerical simulations, and made sure that
the fit is accurate better than 20 percent even at $M =
10^{15}h^{-1}M_\odot$. In any case, the fraction of such massive
clusters is significantly smaller according to equation
(\ref{eq:psmass}), and the difference of the fit and simulations to
that level hardly changes our results in reality.  We should note
here, however, that the bias factor becomes scale-dependent for scales
below $\sim 5\himpc$, and then our assumption of linear and
scale-independent bias breaks down. Thus while our predictions on
linear scales $\simgt 5\himpc$, which we are mostly interested in, are
reliable, those on smaller scales should be interpreted simply to show
the various effects qualitatively.

%%%%%%%%%%%%%%%%%%%%%%%%%%%%%%%%%%%%
\subsection{The light-cone effect}
%%%%%%%%%%%%%%%%%%%%%%%%%%%%%%%%%%%%

The two-point correlation function on the light-cone is properly
formulated by Yamamoto \& Suto (1999). In the present context, their
formula yields the following expression for the two-point correlation
functions of clusters brighter than the X-ray flux-limit $S_{\rm lim}$:
%%%%%%%%%%%%%%%%%%%%%%%%%%%%%%%%%%%%%%%%%%%%%%%%%%%%%%%%%%%%%%%%%%%
\begin{eqnarray}
\label{eq:lcxir1}
    \xi^{\rm LC}_{\rm X-cl}(R; >S_{\rm lim}) 
= {
   {\displaystyle 
     \int_{z_{\rm max}}^{z_{\rm min}} dz 
     {dV_{\rm c} \over dz} ~n_0^2(z)
    \xi^{\rm S}_{\rm cl}(R,z(r);>S_{\rm lim})
    }
\over
    {\displaystyle
     \int_{z_{\rm max}}^{z_{\rm min}} dz 
     {dV_{\rm c} \over dz} ~n_0^2(z)
     }
} 
\end{eqnarray}
%%%%%%%%%%%%%%%%%%%%%%%%%%%%%%%%%%%%%%%%%%%%%%%%%%%%%%%%%%%%%%%%%%%
where $R$ is the comoving separation of a pair of clusters, $z_{\rm
max}$ and $z_{\rm min}$ denote the redshift range of the survey, and
$\xi^{\rm S}_{\rm cl}(R,z;>S_{\rm lim})$ is the corresponding two-point
correlation function on a constant-time hypersurface at $z$ in redshift
space (eq.[\ref{eq:xiScl}]).  The comoving number density of clusters in
the flux-limited survey, $n_0(z; >S_{\rm lim})$, is computed by
integrating the Press -- Schechter mass function as
%%%%%%%%%%%%%%%%%%%%%%%%%%%%%%%%%%%%%%%%%%%%%%%%%%%%%%%%%%%%%%%%%%%
\begin{eqnarray}
\label{eq:n0}
  n_0(z; >S_{\rm lim})
= \int_{M_{\rm lim}(S_{\rm lim})}^\infty n_{\PS}(M,z)dM .
\end{eqnarray}
%%%%%%%%%%%%%%%%%%%%%%%%%%%%%%%%%%%%%%%%%%%%%%%%%%%%%%%%%%%%%%%%%%%
Finally the comoving volume element per unit solid angle is
%%%%%%%%%%%%%%%%%%%%%%%%%%%%%%%%%%%%%%%%%%%%%%%%%%%%%%%%%%%%%%%%%%%
\begin{eqnarray}
\label{eq:dVdz}
{dV_{\rm c} \over dz} = {d_\C^2(z) \over H(z)} .
\end{eqnarray}
%%%%%%%%%%%%%%%%%%%%%%%%%%%%%%%%%%%%%%%%%%%%%%%%%%%%%%%%%%%%%%%%%%%

While equation (\ref{eq:lcxir1}) was originally derived for the
cosmological models with flat spatial geometry (Yamamoto \& Suto 1999),
we assume that the formula is valid for the cosmological models with
open geometry. Note also that this expression for the light-cone effect
looks rather different from that adopted by Mataresse et al. (1997) and
Moscardini et al. (1999a), but that both lead to fairly similar results
quantitatively; see Yamamoto \& Suto (1999) for detailed comparison and
discussion.

The formula (\ref{eq:lcxir1}) is useful in making theoretical
predictions, but $n_0(z)$ itself is not directly observable (unless the
cosmological parameters are specified). Instead it can be rewritten in
terms of the redshift distribution of clusters per unit solid angle with
an X-ray flux $S>S_{\rm lim}$:
%%%%%%%%%%%%%%%%%%%%%%%%%%%%%%%%%%%%%%%%%%%%%%%%%%%%%%%%%%%%%%%%%%%
\begin{eqnarray}
\label{eq:lcxir2}
    \xi^{\rm LC}_{\rm X-cl}(R; >S_{\rm lim}) 
= {
   {\displaystyle 
    \int_\zmin^\zmax {H(z) dz \over d_\C^2(z)}
    \left({dN \over dz}\right)^2 
    \xi^{\rm S}_{\rm cl}(R,z;>S_{\rm lim})
    }
\over
    {\displaystyle
     \int_\zmin^\zmax  {H(z) dz \over d_\C^2(z)}
     \left({dN \over dz}\right)^2 
    }
} ,
\end{eqnarray}
%%%%%%%%%%%%%%%%%%%%%%%%%%%%%%%%%%%%%%%%%%%%%%%%%%%%%%%%%%%%%%%%%%%
and $dN/dz$ which is related to $n_0(z)$ as
%%%%%%%%%%%%%%%%%%%%%%%%%%%%%%%%%%%%%%%%%%%%%%%%%%%%%%%%%%%%%%%%%%%
\begin{eqnarray}
\label{eq:dndz}
{dN \over dz}(z, >S_{\rm lim}) &=& 
  n_0(z; >S_{\rm lim}) ~ {dV_{\rm c} \over dz} .
\end{eqnarray}
%%%%%%%%%%%%%%%%%%%%%%%%%%%%%%%%%%%%%%%%%%%%%%%%%%%%%%%%%%%%%%%%%%%

\begin{figure}
\begin{center}
    \leavevmode\epsfysize=6.75cm \epsfbox{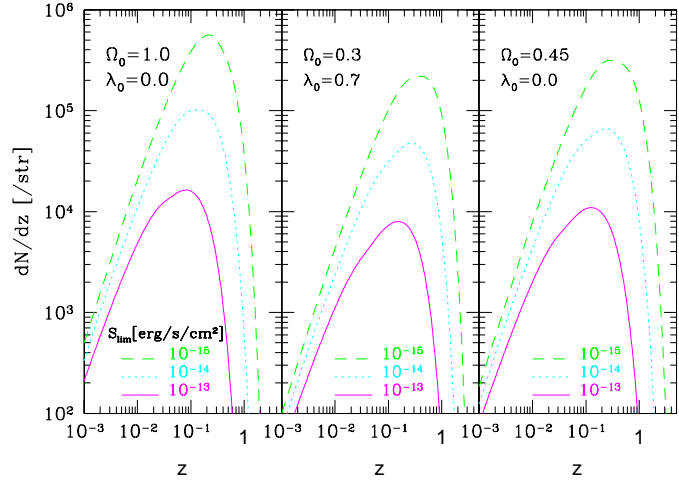}
\end{center}
\figcaption{ Redshift distribution functions for X-ray flux-limited
clusters; $S_{\rm lim}=10^{-13}$, $10^{-14}$ and $10^{-15}$erg/s/cm$^2$
in solid, dotted and dashed lines, respectively. 
\label{fig:dndz}}
\end{figure}

Figure \ref{fig:dndz} plots the redshift distribution function for X-ray
flux-limited clusters with $S_{\rm lim}=10^{-13}$, $10^{-14}$ and
$10^{-15}$ erg/s/cm$^2$. It is interesting to note that $M_{\rm lim}(z)$
corresponding to a fixed $S_{\rm lim}$ {\it decreases} for $z \simgt 1$. 
In fact this behavior is understood from equations (\ref{eq:tm}) to
(\ref{eq:ls}); $M_{\rm lim}(z) \propto T_{\rm lim}^{3/2}(z)/(1+z)^{3/2}
\propto L_{\rm lim}^{3/2\alpha}(z)/(1+z)^{3/2} \propto
d_\L^{3/\alpha}(z)/(1+z)^{3/2}$.

%%%%%%%%%%%%%%%%%%%%%%%%%%%%%%%%%%%%%%%%%%%%%%%%%%%%%%%%%%%%
\section{Predictions for correlations in X-ray flux-limited 
surveys of clusters of galaxies}
%%%%%%%%%%%%%%%%%%%%%%%%%%%%%%%%%%%%%%%%%%%%%%%%%%%%%%%%%%%%%

%%%%%%%%%%%%%%%%%%%%%%%%%%%%%%%%%%%%%%%%%%%%%%%%%%%%%%%%%%%%%%%%
\subsection{Nonlinear evolution, redshift-space distortion and
light-cone effect}
%%%%%%%%%%%%%%%%%%%%%%%%%%%%%%%%%%%%%%%%%%%%%%%%%%%%%%%%%%%%%%%%

\begin{figure}
\begin{center}
    \leavevmode\epsfysize=6.75cm \epsfbox{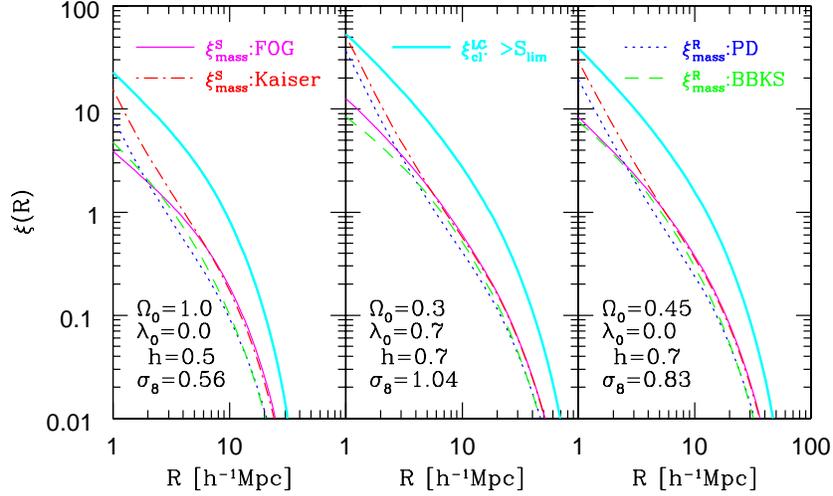}
\end{center}
\figcaption{Light-cone and redshift-space distortion effects on
  two-point correlation functions of clusters; real-space mass
  correlation functions at $z=0$ in linear theory using the BBKS
  transfer function (dashed lines) and in a nonlinear model by PD
  (dashed lines). Redshift-space mass correlation functions at $z=0$
  with the Kaiser distortion (dot-dashed lines) and with the Kaiser
  distortion and finger-of-god (thin solid lines).  Thick solid lines
  indicate our  predictions for the cluster correlation functions
  on the light cone in redshift space (the X-ray flux limit $S_{\rm
    lim}=10^{-14}$erg/s/cm$^2$).
\label{fig:xicl_diff}}
\end{figure}

While our predictions based on the modeling described in the previous
section include various important effects, it would be instructive to
discuss them separately first. For that purpose, we plot in Figure
\ref{fig:xicl_diff} several different predictions for two-point
correlation functions in which some of the effects are artificially
turned off; linear and nonlinear mass correlations in real space at
$z=0$ using the Bardeen et al. (1986; BBKS) and PD formulae for mass
power spectra, cluster correlations with linear redshift-space
distortion (Kaiser 1987) and with full redshift-space distortion, at
$z=0$ using the fitting formula of Mo, Jing \& B\"orner (1997) to
compute $\sigma_v$. These should be compared with our final
predictions on the light-cone in redshift space (eq.[\ref{eq:xiScl}]).

The qualitative features illustrated in Figure \ref{fig:xicl_diff} can
be understood as the combination of the following effects\footnote{As
  mentioned earlier, the behavior of the correlation functions on
  nonlinear scales should not be trusted because we have not properly
  taken into account the possible scale-dependence of the bias below
  the scales.}; the nonlinear gravitational evolution increases the
correlation on small scales, while redshift-space distortion decreases
(enhances) the amplitudes on small (large) scales. The correlation
length defined by eq.[\ref{eq:rc0}] below, for inscance, is enhanced
in redshift space by 50 percent for SCDM and 20 percent for LCDM and
OCDM in comparison with those in real space.  The light-cone effect
averages the amplitude over a range of redshift which is generally
expected to decrease the correlation {\it if} the clustering amplitude
at higher $z$ decreases according to linear theory. In reality,
however, $b_{\rm eff}(z)$ increases more rapidly than the linear
growth rate $D_1(z)$ at higher $z$, and therefore the clustering
amplitude evaluated on the light cone for a given $S_{\rm lim}$
increases as $z$ (see Fig.\ref{fig:rc0_zmax} below).

Nevertheless this is not what we observe in a flux-limited survey.  If
the survey flux-limit $S_{\rm lim}$ becomes smaller, the sample
includes both less luminous clusters at low $z$ and more luminous
clusters at high $z$. While the latter shows stronger bias, the former
should exhibit weaker bias. Thus averaging over the light-cone volume,
their effect on the overall clustering amplitude is fairly
compensated.  This explains why the overall results are not so
sensitive to $S_{\rm lim}$ as shown in left panels of Figure
\ref{fig:xicl_sxtl}.

%%%%%%%%%%%%%%%%%%%%%%%%%%%%%%%%%%%%%%%%%%%%%%%%%%%%%%%%%%%%%%%%%%%%%%%%
\subsection{Effect of the selection function}
%%%%%%%%%%%%%%%%%%%%%%%%%%%%%%%%%%%%%%%%%%%%%%%%%%%%%%%%%%%%%%%%%%%%%%%%

If the temperature of an individual cluster in an X-ray flux-limited
sample is determined observationally, one can construct a
temperature-limited subsample. Similarly one can construct a
luminosity-limited subsample from the flux-limited sample with the
redshift information of each cluster. If the underlying
luminosity--temperature relation (eq. [\ref{eq:lt}]) had no
dispersion, these two subsamples would be essentially the same. In
reality, however, the finite amount of the dispersions would lead to
different predictions for those subsamples, which will provide further
and independent information on the cluster models.  This is
particularly the case here since the bias is highly sensitive to the
mass, and therefore to the temperature and luminosity, of a cluster.

\begin{figure}
\begin{center}
  \leavevmode\epsfxsize=12cm \epsfbox{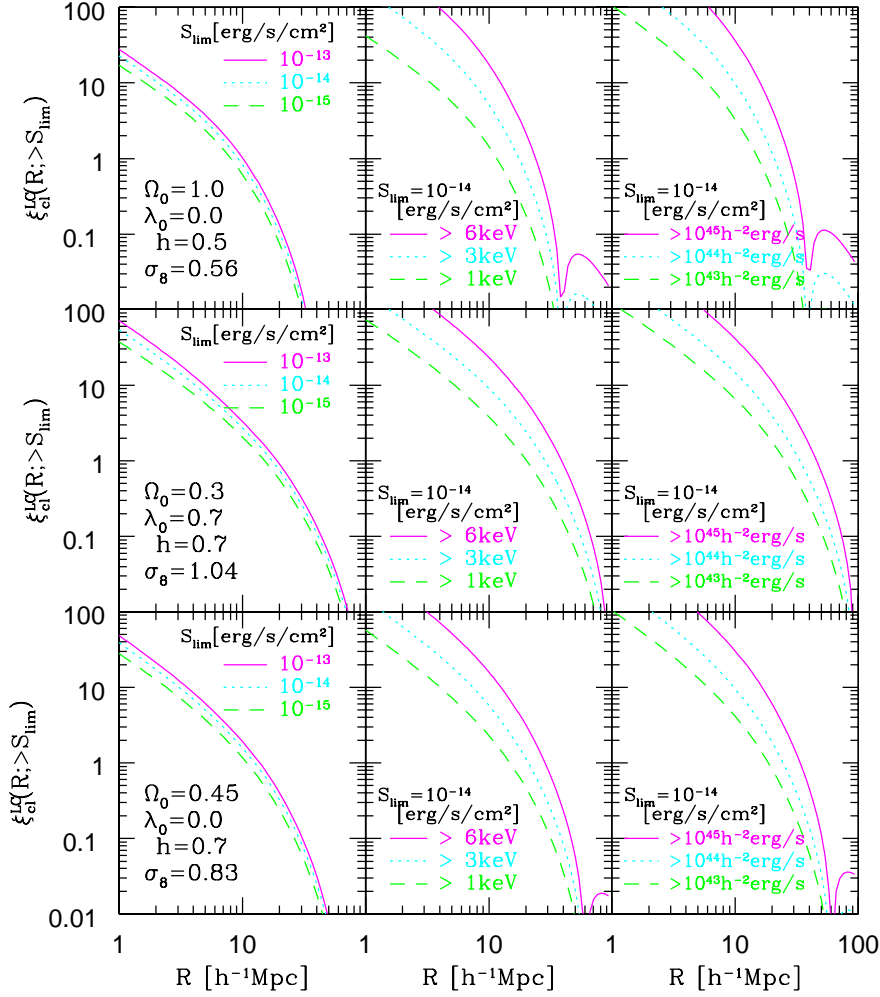} 
\end{center}
\figcaption{Two-point correlation functions of clusters in SCDM ({\it
    Top panels}), LCDM ({\it Middle panels}), and OCDM ({\it Bottom
    panels}) for different selection criteria. {\it Left panels}: the
  X-ray flux limit $S_{\rm lim}=10^{-13}$ (solid lines), $10^{-14}$
  (dotted) and $10^{-15}$erg/s/cm$^2$ (dashed).  {\it Central panels}:
  clusters with the temperature larger than $1$ (solid), $3$ (dotted)
  and $6$keV (dashed) in the X-ray flux-limited sample ($S_{\rm
    lim}=10^{-14}$erg/s/cm$^2$).  {\it Right panels}: clusters with
  the bolometric luminosity larger than $10^{45}$ (solid), $10^{44}$
  (dotted) and $10^{43} h^{-2}$erg/s/cm$^2$ (dashed) in the X-ray
  flux-limited sample ($S_{\rm lim}=10^{-14}$erg/s/cm$^2$).
  \label{fig:xicl_sxtl}}
\end{figure}

Figure \ref{fig:xicl_sxtl} shows our predictions for $\xi^{\rm LC}_{\
cl}(R)$ for cluster samples selected with different flux-limit $S_{\rm
lim}$ ({\it left panels}), and with temperature and absolute
bolometric luminosity limits, $T_{\rm lim}$ and $L_{\rm lim}$ ({\it
middle and right panels}).  For the latter two cases, $S_{\rm
lim}=10^{-14}$erg/s/cm$^2$ is assumed for definiteness.  As explained
in the previous subsection, the results are insensitive to $S_{\rm
lim}$, but very sensitive to $T_{\rm lim}$ and $L_{\rm lim}$,
reflecting the strong dependence of the bias on the latter quantities.

To see this in a somewhat different manner, we plot in Figure
\ref{fig:rc0_sxtl} the correlation length $r_{c0}(S_{\rm lim})$
defined through
%%%%%%%%%%%%%%%%%%%%%%%%%%%%%%%%%%%%%%%%%%%%%%%%%%%% 
\begin{eqnarray}
\label{eq:rc0}
\xi^{\rm LC}_{\rm cl}(r_{c0};>S_{\rm lim})=1,  
\end{eqnarray}
%%%%%%%%%%%%%%%%%%%%%%%%%%%%%%%%%%%%%%%%%%%%%%%%%%%% 
as a function of $S_{\rm lim}$, $T_{\rm lim}$ and $L_{\rm lim}$. For
the latter two cases, $S_{\rm lim}=10^{-14}$erg/s/cm$^2$ is assumed as
in Figure \ref{fig:xicl_sxtl}. Again it is clear that the results are
fairly insensitive to $S_{\rm lim}$, but are sensitive to the bias
factor, which results in the strong dependence of $r_{c0}$ on $T_{\rm
  lim}$ and $L_{\rm lim}$.

The value of $r_{c0}(S_{\rm lim})$ also depends on the depth of the
survey via the light-cone effect and the evolution of bias and mass
fluctuations (Fig.\ref{fig:rc0_zmax}). As a result of the several
competing effects, $r_{c0}(S_{\rm lim})$ increases, albeit very
weakly, as $z_{\rm max}$ becomes larger.

%%%%%%%%%%%%%%%%%%%%%%%%%%%%%%%%%%%%%%%%%%%%%%%%%%%%%%%%%%%%%%%%%%%%%%%%%%
\subsection{Dependence on $\Omega_0$}
%%%%%%%%%%%%%%%%%%%%%%%%%%%%%%%%%%%%%%%%%%%%%%%%%%%%%%%%%%%%%%%%%%%%%%%%%%

For a cosmological application of the present result, it is
interesting to examine how the $r_{c,0}(S_{\rm lim})$ depends on
$\Omega_0$.  This is summarized in Figure \ref{fig:rc0_omega0}, where
we fix the value of the fluctuation amplitude $\sigma_8$ adopting the
cluster abundance constraint (eq.[\ref{eq:s8o0}]), and consider both
$\lambda_0=1-\Omega_0$ (thick lines) and $\lambda_0=0$ (thin lines).
We set the shape parameter of the spectrum $\Gamma$ as $\Omega_0 h
\exp[-\Omega_{\rm b}(1+\sqrt{2 h}\Omega_0^{-1})]$ with $\Omega_{\rm
  b}h^2=0.015$ and $h=0.7$.  If we adopt such spectra, the correlation
length in $\lambda_0=1-\Omega_0$ model is generally larger than that
in $\lambda_0=0$ model, largely reflecting the dependence of
$\sigma_8$ on $\lambda_0$ (eq.[\ref{eq:s8o0}]).  Again the results are
not sensitive to the flux limit $S_{\rm lim}$.  The dependence on
$\Omega_0$ is rather strong, and these predictions combined with the
future observational results will be able to break the degeneracy of
the cosmological parameters.

%%%%%%%%%%%%%%%%%%%%%%%%%%%%%%%%%%%%%%
\section{Conclusions and discussion}
%%%%%%%%%%%%%%%%%%%%%%%%%%%%%%%%%%%%%%

We have presented a detailed methodology to predict the two-point
correlation functions for X-ray flux-limited samples of clusters of
galaxies, fully taking into account the redshift-space distortion,
nonlinear gravitational evolution of mass fluctuations, evolution of
bias, the light-cone effect and the observational selection function.
While our method is similar to the recent model of Moscardini et al.
(1999a) in many respects, the most important difference is that we
incorporated both linear and nonlinear redshift-space distortion
following the prescription of Suto et al. (1999), Yamamoto, Nishioka
\& Suto (1999) and Magira, Jing \& Suto (2000).

The predictions for correlations of clusters are most sensitive to the
model of bias among others.  Fortunately, the Press -- Schechter theory
provides a quite reliable bias model for clusters (Mo \& White 1996;
Jing 1998), which is in marked contrast with a model for galaxy and
quasar bias. Thus one can make quantitative and detailed model
predictions once a set of the cosmological parameters are specified.
These predictions can be checked against the flux-limited samples of
clusters from the future X-ray satellites including Astro-E, Chandra,
and XMM. It is particularly interesting to probe the value of $\Omega_0$
and the degree of non-Gaussianity by combining this comparison and the
cluster abundance (Kitayama, Sasaki, \& Suto 1998; Koyama, Soda, \&
Taruya 1999).

We should note, however, that the current methodology is reliable
only for linear scales; on small scales ($\simlt 5 \himpc$), the
approximation of the linear and scale-independent bias breaks down. In
addition, the projection effect becomes inevitably important below a
few \himpc scales even for the X-ray selected samples, and the
observational data analysis itself becomes non-trivial.

While the selection function for the future survey would not exactly
match those in our examples, it is straightforward to make suitable
predictions on the basis of the present formalism at least on linear
scales, and we hope that we have already presented all the basic and
qualitative features of the clustering statistics.

\begin{figure}
\begin{center}
  \leavevmode\epsfxsize=12cm \epsfbox{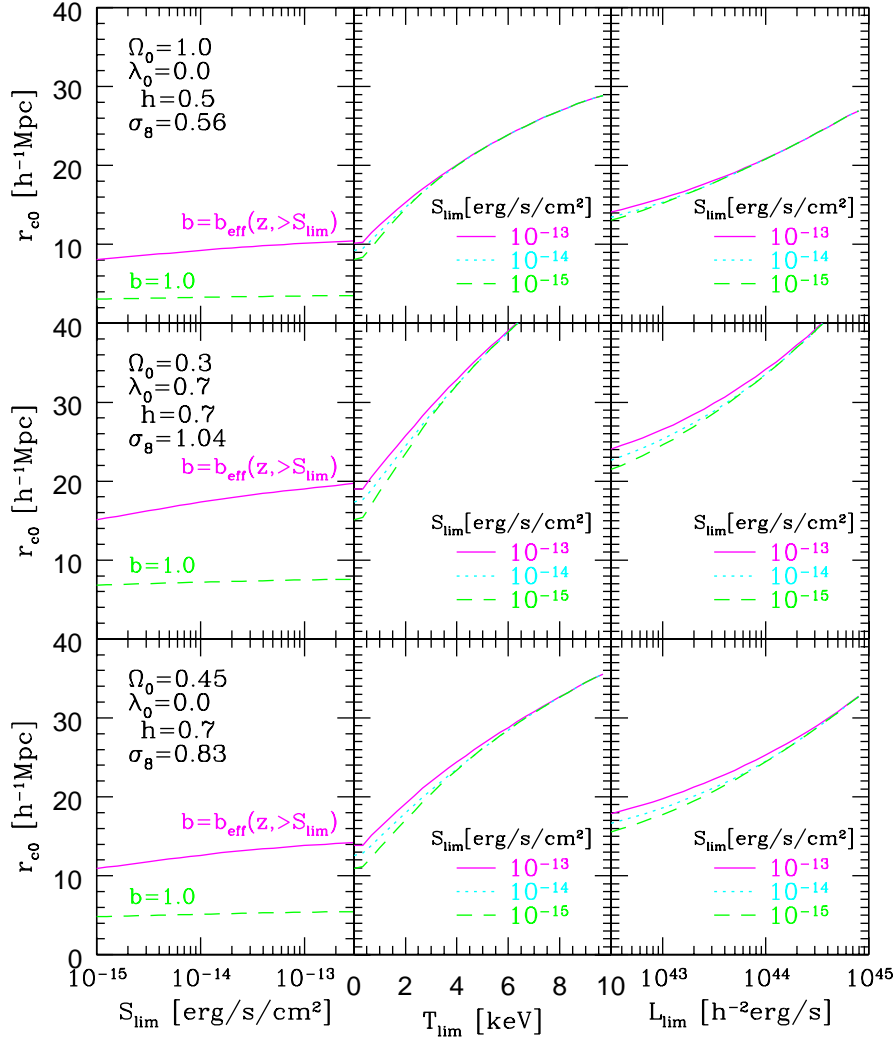} 
\end{center}
\figcaption{Correlation lengths of clusters $r_{\rm c0}$ in SCDM ({\it
    Top panels}), LCDM ({\it Middle panels}), and OCDM ({\it Bottom
    panels}) for different selection criteria. {\it Left panels}:
    $r_{\rm c0}$ for the X-ray flux-limited clusters.  Dashed lines
    refer to the results in which the bias parameter is set to unity,
    while solid lines indicate our predictions.  {\it Central panels}:
    $r_{\rm c0}$ as a function of the temperature limit for the X-ray
    flux-limited sample of $S_{\rm lim}=10^{-13}$ (solid lines),
    $10^{-14}$ (dotted) and $10^{-15}$erg/s/cm$^2$ (dashed). {\it
    Right panels}: $r_{\rm c0}$ as a function of the bolometric
    luminosity limit for the X-ray flux-limited sample of $S_{\rm
    lim}=10^{-13}$ (solid lines), $10^{-14}$ (dotted) and
    $10^{-15}$erg/s/cm$^2$ (dashed).
\label{fig:rc0_sxtl}} 
\end{figure}

After submitting this paper, Moscardini et al. (1999b) posted a quite
similar paper on the clustering of X-ray selected galaxy clusters.
They predicted the correlation lengths at $S_{\rm lim} =
10^{-14}$erg/s/cm$^2$ of 7 and 12 \himpc in SCDM and LCDM models,
respectively (their Fig.7), which should be compared with our
predictions of 9 and 17 \himpc (Fig.\ref{fig:rc0_sxtl}). While our
models adopt slightly different values for the shape parameter
$\Gamma$, and the normalization of fluctuations, $\sigma_8$, it is
clear that our correlation lengths are systematically larger by $30
\sim 40$ percent level. This should be ascribed to be the
redshift-distortion effect which they neglect in their analysis.  In
fact, this difference is comparable to the cosmological model
dependence which they showed in their Figure 7, clearly suggesting the
importance of the effect to this level. Apart from this difference,
however, our results are consistent with each other.

\bigskip
\bigskip

We thank an anonymous referee for detailed comments which improved the
presentation of the earlier manuscript.  Y.P.J. and T.K. gratefully
acknowledge the fellowship from the Japan Society for the Promotion of
Science.  Part of numerical computations was carried out on VPP300/16R
and VX/4R at ADAC (the Astronomical Data Analysis Center) of the
National Astronomical Observatory, Japan, as well as at RESCEU
(Research Center for the Early Universe, University of Tokyo) and KEK
(High Energy Accelerator Research Organization, Japan).  This research
was supported in part by the Grants-in-Aid by the Ministry of
Education, Science, Sports and Culture of Japan to RESCEU (07CE2002)
and to K.Y. (11640280), and by the Inamori Foundation.

\newpage
%%%%%%%%%%%%%%%%%%%%%%%%%%%%%%%%%%%%%%%%%%%%%%%%%%%%%%%%%%%%%%%%%%%%
\parskip4pt
\baselineskip 13pt

\centerline{\bf REFERENCES}
\bigskip

\def\apjpap#1;#2;#3;#4; {\pp#1, {#2}, {#3}, #4}
\def\apjbook#1;#2;#3;#4; {\pp#1, {#2} (#3: #4)}
\def\apjppt#1;#2; {\pp#1, #2.}
\def\apjproc#1;#2;#3;#4;#5;#6; {\pp#1, {#2} #3, (#4: #5), #6}

\apjpap Bahcall, N.A. \& Soneira, R.M. 1983;ApJ;270;20;
\apjpap Bahcall, N.A. 1988;ARA\&A;26;631;
\apjpap Bahcall, N.A. \& Cen, R.Y. 1993;ApJ;407;L49;
\apjpap Barbosa, D., Bartlett, J. G., Blanchard, A., \& Oukbir, J. 1996;
     AA;314;13;
\apjpap Bardeen, J. M., Bond, J. R., Kaiser, N., \& Szalay, A. S. 1985
      ;ApJ;304;15 (BBKS);
\apjpap Borgani, S., Plionis, M., \& Kolokotronis, V. 1999;MNRAS;305;866;
\apjpap Burns, J.F. et al. 1996;ApJ;467;49;
\apjpap Cole, S., Fisher, K. B., \& Weinberg, D. H. 1994;MNRAS;267;785;
\apjpap Cole, S., Fisher, K. B., \& Weinberg, D. H. 1995;MNRAS;275;515;
\apjpap David, L. P., Slyz, A., Jones, C., Forman, W., \& Vrtilek, S. D. 
   1993;ApJ;412;479; 
\apjpap Ebeling, H., Edge, A. C., Fabian, A.C., Allen, S. W., 
   \& Crawford C. S. 1997;ApJ;479;L101;
\apjpap Ebeling, H., Voges, W., B\"{o}hringer, H., Edge, A. C.,
  Huchra, J. P., \& Briel, U. G. 1996;MNRAS;281;799;
\apjpap Eke, V. R., Cole, S., \& Frenk, C. S. 1996;MNRAS;282;263;
\apjppt Hamilton, A. J. S. 1998; in `` The Evolving Universe. Selected
Topics on Large-Scale Structure and on the Properties of Galaxies'',
(Kluwer: Dordrecht), p.185;
\apjpap Jing, Y. P. 1998;ApJ;503;L9;
\apjpap Jing, Y.P., Fang, L. Z. 1994; ApJ; 432; 438;
\apjpap Jing, Y.P., Mo H.J., B\"orner, G., Fang, L.Z. 1993; ApJ; 411; 450;
\apjpap Kaiser, N. 1984;ApJ;284;L9;
\apjpap Kaiser, N. 1987;MNRAS;227;1;
\apjpap Kitayama, T., Sasaki,S., \& Suto, Y. 1998;PASJ;50;1;
\apjpap Kitayama, T., \& Suto, Y. 1996;ApJ;469;480;
\apjpap Kitayama, T., \& Suto, Y. 1997;ApJ;490;557;
\apjpap Klypin, A.A. \& Kopylov, A.I., 1983;Sov.Astron.Lett.;9;75;
\apjppt  Koyama, K.,  Soda, J., \& Taruya, A. 1999;
  MNRAS, in press (astro-ph/9903027);
\apjpap Magira, H., Jing, Y. P., \& Suto, Y. 2000;ApJ;
528;January 1st issue, in press (astro-ph/9907438);
\apjpap Makino, N., Sasaki, S., \& Suto, Y. 1998;ApJ;497;555;
\apjpap Masai, K. 1984;Ap\&SS;98;367;
\apjpap Matarrese, S., Coles, P., Lucchin, F., \& Moscardini, L. 1997;
 MNRAS;286;115;
\apjpap Matsubara, T., \& Suto, Y. 1996;ApJ;470;L1;
\apjpap Matsubara, T., Suto, Y., \& Szapudi, I. 1997;ApJ;491;L1;
\apjpap Mo, H. J., Jing, Y. P., \& B\"orner, G. 1997;MNRAS;286;979;
\apjpap Mo, H. J., \& White, S. D. M. 1996;MNRAS;282;347;
\apjpap Moscardini, L., Coles, P., Lucchin, F., \& Matarrese, S. 1998
   ;MNRAS;299;95;
\apjppt Moscardini, L., Matarrese, S., De Grandi, S. \&  Lucchin, F. 1999a;
   MNRAS, submitted (astro-ph/9904282);
\apjppt Moscardini, L., Matarrese, S., Lucchin, F., \& Rosati, P. 1999b;
   MNRAS, submitted (astro-ph/9909273);
\apjppt Mushotzky, R. F., \& Scharf, C. A. 1997;ApJ;482;L13;
\apjpap Nakamura, T. T., Matsubara, T., \& Suto, Y. 1998;ApJ;494;13;
\apjpap Nishioka, H., \& Yamamoto, K. 1999;ApJ;520;426;
\apjpap Peacock, J. A., \& Dodds, S. J. 1996;MNRAS;280;L19 (PD);
\apjpap Ponman, T. J., Bourner, P. D. J., Ebeling, H., \&
  B\"{o}hringer, H. 1996;MNRAS;283;690;
\apjpap Rosati, P., Della Ceca, R., Norman, C., \& Giacconi, R. 1998;
 ApJ;492;L21;
\apjpap Suto, Y., Magira, H., Jing, Y. P., Matsubara, T., 
 \& Yamamoto, K. 1999;Prog.Theor.Phys.Suppl.;133;183;
\apjpap Suto, Y., Sasaki, S. \& Makino, N. 1998;ApJ;509;544;
\apjpap Ueda, H., Itoh, M., \& Suto, Y. 1993;ApJ;408;3;
\apjpap Viana, P. T. P., \& Liddle, A. R. 1996;MNRAS;281;323;
\apjpap Watanabe, T., Matsubara, T., \& Suto, Y. 1994;ApJ;432;17;
\apjpap White, S. D. M., Efstathiou, G., \& Frenk, C. S. 1993;
   MNRAS;262;1023;
\apjpap Yamamoto, K., Nishioka, H., \& Suto, Y. 1999;ApJ;
527;December 20 issue, in press (astro-ph/9908006); 
\apjpap Yamamoto, K., \& Suto, Y. 1999;ApJ;517;1;
\apjpap Yoshikawa, K., Itoh, M., \& Suto, Y. 1998;PASJ;50;203;
\apjppt Yoshikawa, K., Jing, Y.P., \& Suto, Y. 1999;ApJ, submitted;
\apjpap Yoshikawa, K., \& Suto, Y. 1999;ApJ;513;549;

\clearpage

%% F1 %%%%%%%%%%%%%%%%%%%%%%%%%%%%%%%%%%%%%%%%%%%%%%%%%%%%%%%%%%%%%%%%%%
%%%%%%%%%%%%%%%%%%%%%%%%%%%%%%%%%%%%%%%%%%%%%%%%%%%%%%%%%%%%%%%%%%%%%

%%% F2 %%%%%%%%%%%%%%%%%%%%%%%%%%%%%%%%%%%%%%%%%%%%%%%%%%%%%%%%%%%%%%%
%%%%%%%%%%%%%%%%%%%%%%%%%%%%%%%%%%%%%%%%%%%%%%%%%%%%%%%%%%%%%%%%%%%%%

%%% F3 %%%%%%%%%%%%%%%%%%%%%%%%%%%%%%%%%%%%%%%%%%%%%%%%%%%%%%%%%%%%%%
%%%%%%%%%%%%%%%%%%%%%%%%%%%%%%%%%%%%%%%%%%%%%%%%%%%%%%%%%%%%%%%%%%%%%

%%% F4 %%%%%%%%%%%%%%%%%%%%%%%%%%%%%%%%%%%%%%%%%%%%%%%%%%%%%%%%%%%%%%
%%%%%%%%%%%%%%%%%%%%%%%%%%%%%%%%%%%%%%%%%%%%%%%%%%%%%%%%%%%%%%%%%%%%%

%%% F5 %%%%%%%%%%%%%%%%%%%%%%%%%%%%%%%%%%%%%%%%%%%%%%%%%%%%%%%%%%%%%%
%%%%%%%%%%%%%%%%%%%%%%%%%%%%%%%%%%%%%%%%%%%%%%%%%%%%%%%%%%%%%%%%%%%%%

%%% F6 %%%%%%%%%%%%%%%%%%%%%%%%%%%%%%%%%%%%%%%%%%%%%%%%%%%%%%%%%%%%%%
%%%%%%%%%%%%%%%%%%%%%%%%%%%%%%%%%%%%%%%%%%%%%%%%%%%%%%%%%%%%%%%%%%%%%

%%% F7 %%%%%%%%%%%%%%%%%%%%%%%%%%%%%%%%%%%%%%%%%%%%%%%%%%%%%%%%%%%%%
%%%%%%%%%%%%%%%%%%%%%%%%%%%%%%%%%%%%%%%%%%%%%%%%%%%%%%%%%%%%%%%%%%%%%

%%% F8 %%%%%%%%%%%%%%%%%%%%%%%%%%%%%%%%%%%%%%%%%%%%%%%%%%%%%%%%%%%%%%
%%%%%%%%%%%%%%%%%%%%%%%%%%%%%%%%%%%%%%%%%%%%%%%%%%%%%%%%%%%%%%%%%%%%%

%%% F9 %%%%%%%%%%%%%%%%%%%%%%%%%%%%%%%%%%%%%%%%%%%%%%%%%%%%%%%%%%%%%%
\begin{figure}
\begin{center}
  \leavevmode\epsfysize=6.75cm \epsfbox{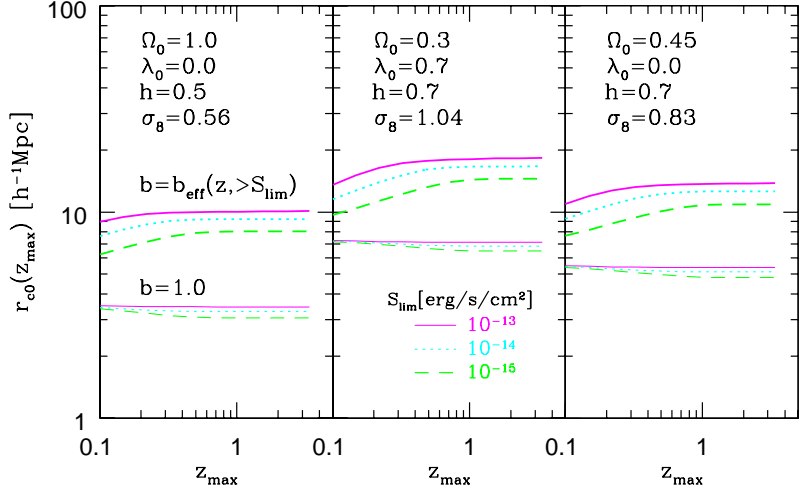} 
\end{center}
\figcaption{Correlation lengths of clusters in SCDM ({\it Left
    panels}), LCDM ({\it Middle panels}), and OCDM ({\it Right
    panels}) as a function of the survey depth, $z_{\rm max}$. The
  X-ray flux-limit $S_{\rm lim}$ is $10^{-13}$ (solid), $10^{-14}$
  (dotted) and $10^{-15}$erg/s/cm$^2$ (dashed lines).  Thin lines
  indicate the predictions in the $b(z)=1$ model for comparison.
\label{fig:rc0_zmax}}
\end{figure}
%%%%%%%%%%%%%%%%%%%%%%%%%%%%%%%%%%%%%%%%%%%%%%%%%%%%%%%%%%%%%%%%%%%%%

%%% F10 %%%%%%%%%%%%%%%%%%%%%%%%%%%%%%%%%%%%%%%%%%%%%%%%%%%%%%%%%%%%%%
\begin{figure}
\begin{center}
  \leavevmode\epsfysize=6.75cm \epsfbox{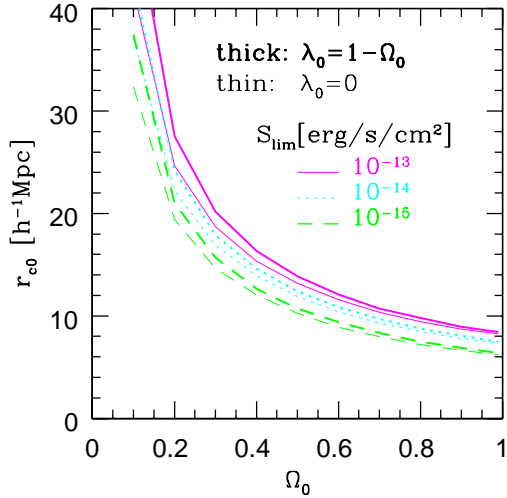} 
\end{center}
\figcaption{Correlation lengths of clusters as a function of
  $\Omega_0$. The shape parameter of the spectrum $\Gamma$ is fixed as
  $\Omega_0 h \exp[-\Omega_{\rm b}(1+\sqrt{2 h}\Omega_0^{-1})]$ with
  $\Omega_{\rm b}h^2=0.015$ and $h=0.7$.  The X-ray flux-limit $S_{\rm
    lim}$ is $10^{-13}$ (solid lines), $10^{-14}$ (dotted) and
  $10^{-15}$erg/s/cm$^2$ (dashed). For each $S_{\rm lim}$, we plot the
  case of $\lambda_0=1-\Omega_0$ in thick lines, and $\lambda_0=0$ in
  thin lines.
\label{fig:rc0_omega0}} 
\end{figure}
%%%%%%%%%%%%%%%%%%%%%%%%%%%%%%%%%%%%%%%%%%%%%%%%%%%%%%%%%%%%%%%%%%%%%

\end{document}